\documentclass{article}
\usepackage{arxiv}
\usepackage{times}
\usepackage{ifpdf}
\usepackage[english]{babel}
\usepackage{cite}
\usepackage{tabularx}
\usepackage{microtype}
\usepackage{multirow}
\usepackage{amsmath}
\usepackage{lipsum}
\usepackage{comment}
\usepackage{hyperref}
\usepackage[caption=false, font=footnotesize]{subfig}% Modern replacement for subfigure package
\usepackage{paralist}% extended list environments
\usepackage[figure,table]{hypcap}% hyperref companion
\usepackage[english]{babel}
\usepackage[]{graphicx}
\def\papertitle{Real-time Timbre Transfer and Sound Synthesis using DDSP}

\author{Francesco Ganis, Erik Frej Knudesn, Søren V.~K. Lyster, Robin Otterbein, David Südholt and Cumhur Erkut\\
  Department of Architecture, Design, and Media Technology\\
  Aalborg University Copenhagen, Denmark\\
  \url{https://www.smc.aau.dk/}}

\title{\papertitle}
\begin{document}
\maketitle
\begin{abstract}
Neural audio synthesis is an actively researched topic, having yielded a wide range of techniques that leverages machine learning architectures. Google Magenta elaborated a novel approach called Differential Digital Signal Processing (DDSP) that incorporates deep neural networks with preconditioned digital signal processing techniques, reaching state-of-the-art results especially in timbre transfer applications. However, most of these techniques, including the DDSP, are generally not applicable in real-time constraints, making them ineligible in a musical workflow. In this paper, we present a real-time implementation of the DDSP library embedded in a virtual synthesizer as a plug-in that can be used in a Digital Audio Workstation. We focused on timbre transfer from learned representations of real instruments to arbitrary sound inputs as well as controlling these models by MIDI. Furthermore, we developed a GUI for intuitive high-level controls which can be used for post-processing and manipulating the parameters estimated by the neural network. We have conducted a user experience test with seven participants online. The results indicated that our users found the  interface appealing, easy to understand, and worth exploring further. At the same time, we have identified issues in the timbre transfer quality, in some components we did not implement, and in installation and distribution of our plugin. The next iteration of our design will address these issues. Our real-time MATLAB and JUCE implementations are available at \url{https://github.com/SMC704/juce-ddsp} and \url{https://github.com/SMC704/matlab-ddsp}, respectively.
\end{abstract}

\section{Introduction}\label{sec:introduction}
Sound synthesizers have been widely used in music production since the late 50s. Because of their inner complexity, many musicians and producers polish presets' parameters until they reach the desired sound. This procedure is time-consuming and sometimes results in failed attempts to achieve a desired sound.

Much research has been done in the area of automating the generation of these sounds through the aid of machine learning and neural networks. Common approaches included directly generating the waveform in the time domain \cite{donahue_adversarial_2019} or predicting synthesis parameters based on hand-picked analysis features \cite{Blaauw2017}. In their 2020 paper on Differentiable Digital Signal Processing (DDSP)\cite{engel_ddsp_2020}, Engel et al.\ proposed a novel approach to neural audio synthesis. Rather than generating signals directly in the time or frequency domain, DDSP offers a complete end-to-end toolbox consisting of a synthesizer based on Spectral Modeling Synthesis (SMS) \cite{serra_spectral_1990}, and an autoencoder neural network architecture that takes care of both extracting analysis features and predicting synthesis parameters.

% The SMS sound generation combines an additive synthesizer, parameterized by the frequencies and amplitudes of many sinusoids, and a subtractive synthesizer, parameterized by the magnitude response of a linear filter that is applied to white noise. We will refer to this combined synthesizer as the \textit{back-end}.

%The \textit{autoencoder} network consists of an encoder trained to extract loudness and pitch (and optionally timbre information) and a decoder trained to map the encoder output to parameters for the back-end. A model trained on recordings of e.g. a violin could now extract loudness and pitch information from any arbitrary audio input, and generate a sequence of parameters for the back-end that synthesize the sound of a violin playing at that same loudness and pitch. % TODO this could probably use some more elegant wordings & stuff
The authors of the DDSP paper released a public demonstration of "tone transfer"\footnote{\url{https://sites.research.google/tonetransfer}, last accessed on 2020-11-30}, allowing the user to upload their own recordings, select from a list of models trained on various instruments and "transfer" their recorded melodies to the sound of a trumpet, a violin etc. % This demonstration reached an audience far beyond the research community when it was showcased by several popular music-focused YouTube channels \cite{neely_2020, huang_2020}.

% This project aims to 

\begin{figure}[bthp]
\centering
\includegraphics[width=0.99\columnwidth]{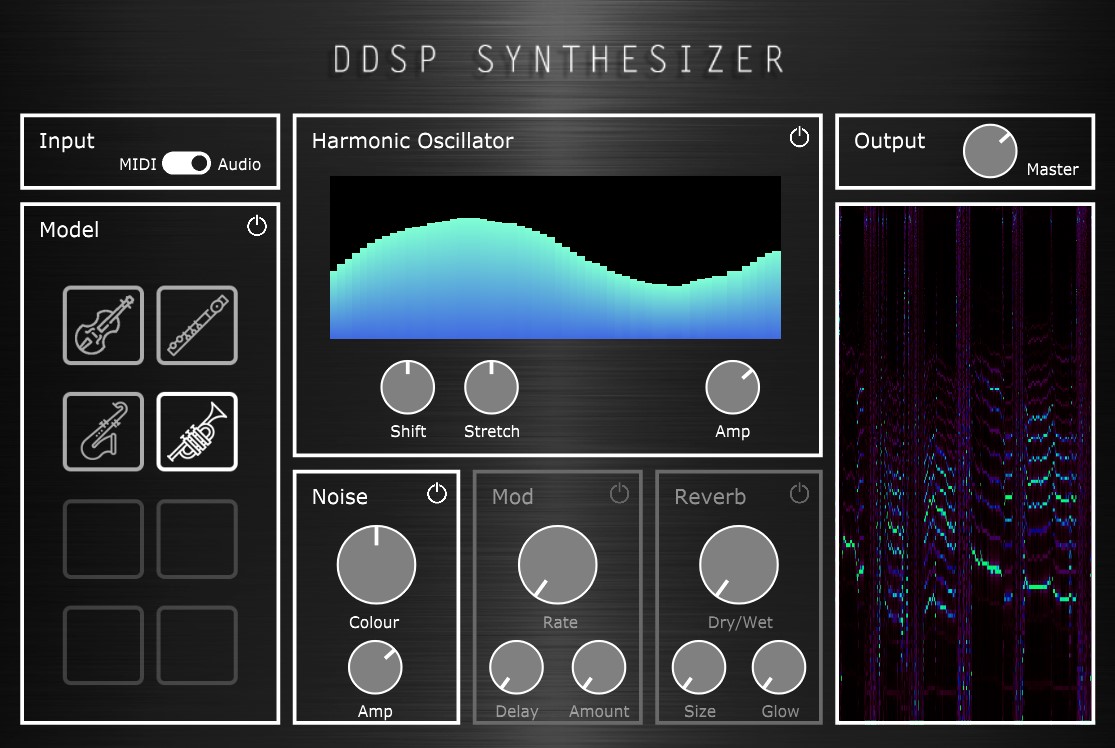}
\caption{Our real-time DDSP Synthesizer GUI.}
\label{fig:ddsp_gui}
\end{figure}

We implemented the DDSP back-end as a virtual instrument playable in real-time. Figure \ref{fig:ddsp_gui} shows the GUI of our synthesizer. This paper documents the background, our requirement-driven design and implementation approach, including model compenents and training, the GUI design, and user experience evaluation. The structure of this paper follows these main topics in order. 

Besides our contribution to the real-time neural audio synthesis, we release our real-time MATLAB and JUCE implementations at \url{https://github.com/SMC704/juce-ddsp} and \url{https://github.com/SMC704/matlab-ddsp}, respectively.
  % TODO some more stuff for motivation, models etc

\section{Background}\label{sec:background}
%In this section we will briefly describe the main components of DDSP library and the research path conducted by the authors during the initial stages of this project.

In addition to the DDSP paper \cite{engel_ddsp_2020}, our work is inspired by the commercially produced additive synthesizer called \emph{Razor} by Native Instruments\cite{native-instruments}. Razor's core consists of a powerful additive synthesizer and features various modulation options for manipulating the sound output. What is especially interesting about Razor is that every modulation option (e.g. filters, stereo imaging, reverbs and delays) is actually modulating individual partial harmonics (non-integer multiples of the fundamental frequency) in the additive synthesis engine. Furthermore, Razor enables musicians and producers to intuitively control partials via different parameters while relying on a visual representation of partial manipulation. We therefore focused on the harmonic and the stochastic components of the DDSP.
%
%\subsection{DDSP library}\label{subsec:DDSP_library}
%In the next paragraphs we will summarize the theory behind the synthesis core of the DDSP library. Here we report an excerpt from the original publication \cite{engel_ddsp_2020}.  

% The main idea behind the Spectral Model Synthesis technique is to emulate the natural sounds recreating both 
 %of the sounds,  \cite{serra_spectral_1990}, specifically % Thus, this algorithm combines 
 %to obtain remarkably realistic results.

\begin{comment}
\begin{figure}[htp]
\centering
\includegraphics[width=0.9\columnwidth]{SMC704 DDSP/Images/ddsp_autoencoder.png}
\caption{Autoencoder architecture from the DDSP Library \cite{engel_ddsp_2020}
\label{fig:ddsp_autoencoder}}
\end{figure}
\end{comment}

\subsection{Harmonic Oscillator / Additive Synthesizer}\label{subsubsec:additive_synth}
The additive synthesizer is the main core of the whole synthesis and is responsible for generating all the harmonic components of the reconstructed sound. The output is characterized by the sum of several harmonic integer multiples of the fundamental frequency $f_0$:

\begin{equation}
f_k = k \cdot f_0(n) .
\label{eq:kth_harmonic}
\end {equation}

In order to generate the harmonics, we can implement $k$ oscillators in the discrete time:

\begin{equation}
x(n) = \sum_{k = 1}^{K} A_k(n) \cdot sin(\phi_k(n)), 
\label{eq:add_synth}
\end {equation} where $A_k(n)$ is the time-varying amplitude of the $k_{th}$ sinusoidal component and $\phi_k(n)$ is its instantaneous phase. $\phi_k(n)$ is obtained using equation \ref{eq:phase}.

\begin{equation}
\phi_k(n) = 2\pi \sum_{m = 0}^{n} f_k(m) + \phi_{0,k}.
\label{eq:phase}
\end {equation}

The only two parameters necessary to control the synthesizer are the frequency $f_0(n)$ and the harmonic amplitudes $A_k(n)$. These are retrieved directly from the input sound using the encoder contained in the autoencoder network. As reported in \cite{engel_ddsp_2020}, the network outputs are scaled and normalized to fall within an interpretable value range for the synthesizer

\subsection{Filtered Noise / Subtractive Synthesizer}\label{subsubsec:ddsp_subtractive_synth}
The subtractive synthesis is used to recreate the non-harmonic part of natural sounds. The parameters necessary to obtain a frequency-domain transfer function of a linear time-variant finite impulse response (LTV-FIR) filter are retrieved from the neural network in frames that are subsets of the input signal. The corresponding impulse responses (IRs) are calculated and a windowing function is applied. The windowed IRs are then convolved with white noise via transformation to and multiplication in the frequency domain.

\subsection{Reverb}\label{subsubsec:reverb}
In addition to the SMS model, the sound is also given a sense of space using a reverberation algorithm performed in the frequency domain. Thus, the operation of convolution between the impulse response of the reverb and the synthesized signal is a more efficient multiplication.

\subsection{Research question}\label{subsec:research_question_statement}
Based on this background % problem exploration in \ref{subsec:exploration_problem_space} we have narrowed down the scope of our project by 
we have formulated the following research question: How can we develop a playable software instrument, based on the DDSP library, that would: a) allow customization of model-estimated synth parameters through top-level macro controls, b) enable existing workflow-integration in Digital Audio Workstations (DAWs), and c) facilitate a simple approach for beginners without limiting usability for expert music producers?

To sum up the design requirements, we aim to build a software instrument plugin that is playable in real-time. The instrument must support different composition techniques, thus having a line and MIDI input mode. The instrument must include at least four pre-trained models which serve the purpose of estimating synthesizer parameters to output a desired sound. Finally, the instrument must include graphical user interface components that provide intuitive controls for the manipulation of synthesizer and effect parameters.

\section{Design \& Implementation}\label{sec:implementation}
Based on this research question, we have identified five user needs \cite{Pandey2010}, and matched them with a solution, reformulating them as a concrete measurable design requirement. The design requirements are thus documented on Table \ref{tab:design_requirements}.

\begin{table*}[ht]
    \begin{center}
        \begin{tabularx}{0.99\columnwidth}{|l|X|X|X|}
            \hline
             \textbf{\#} & \textbf{User Obj.} & \textbf{Solution} & \textbf{Design Requirement} \\ 
             \hline
             1 & Provide a new playable instrument for unique sound generation and inspiration & Real-time implementation & \emph{Must work in real-time as a playable software instrument.}  \\ \hline
             2 & Conveniently integrate into existing workflows & Plugin format application & \emph{Must be implemented as a software plugin.}\\ 
             \hline
             3 & Adapt to different composition methods & Allow line and MIDI input & \emph{Must allow switching between Line and MIDI input.} \\ 
             \hline
             4 & Easy fast unique sound generation & Choose models for sound generation & \emph{Must implement at least four pre-trained models.} \\ 
             \hline
             5 & Convenient customizability of sounds & Tweakable parameters that effects the audio output & \emph{Must include GUI components for intuitive manipulation of synth and effects parameters.} \\ \hline
        \end{tabularx}
    \end{center}
    \caption{Documentation of Design Requirements}
    \label{tab:design_requirements}
\end{table*}

%Technical aspects of solution

\subsection{Architecture overview}\label{subsec:architecture_overview}

\begin{figure*}[htp]
\centering
\includegraphics[width=0.99\columnwidth]{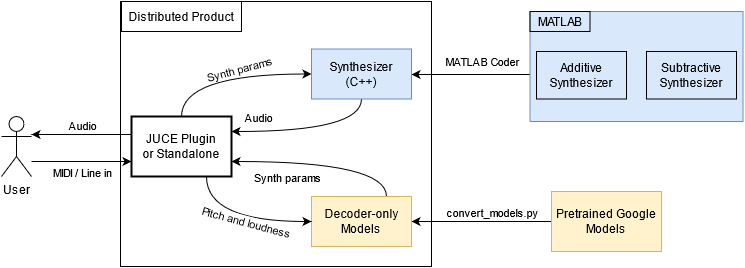}
\caption{Schematic overview of the project architecture.
\label{fig:architecture_scheme}}
\end{figure*}

To meet our criteria of creating a real-time software instrument, we decided to build the plugin in C++ using the JUCE application framework\footnote{\url{https://juce.com/}, last accessed on 2020-12-15}. With JUCE, we had a multi-platform supported audio plugin template that was handling MIDI and audio inputs and outputs. This allowed us to mainly focus on the audio processing and GUI. 

Creating a real-time implementation of the non-real-time DDSP library posed some immediate challenges. To analyze and understand these challenges we decided to start by doing a direct translation of the additive and subtractive synthesizers from the DDSP library into MATLAB. The synthesizers could then be changed into real-time implementations and tested. In order to use our MATLAB implementation in the JUCE framework, we used inbuilt MATLAB tools to generate C++ code. 

We transformed the autoencoder models pretrained by Google into models that could be used to estimate synthesizer parameters directly from our plugin's user input.

% The JUCE implementation of this project can be found here\footnote{\url{https://github.com/SMC704/juce-ddsp}, last accessed on 2020-12-18}, and the MATLAB implementation can be found here\footnote{\url{https://github.com/SMC704/matlab-ddsp}, last accessed on 2020-12-18}. 

A general overview of this architecture can be seen in figure \ref{fig:architecture_scheme}.
The following sections will discuss each component in more detail.

\subsubsection{Synth in MATLAB}\label{subsubsec:synth_MATLAB}
MATLAB’s environment and visualization tools gave us access to quick prototyping and testing. This allowed us to do the implementation over multiple iterations. We tested our synthesizers' compatibility with the predicted parameters from the DDSP models by invoking the encoders and decoders in isolation through MATLAB's Python interface.

At first we implemented the non-real-time synthesis algorithms of the DDSP library. Then the synthesizers were changed to real-time, i.e., synthesizing a single frame at a time.   
Using the MATLAB Audio Test Bench, we could then test the functionality of the synthesizer components and parameters with real-time audio and varying sample rate and buffer size. 
The last iterations consisted of optimizing the code with the constraints of real-time audio processing on CPUs.  

\subsubsection{MATLAB to C++}\label{subsubsec:MATLAB_to_C++}
Using the MATLAB coder tool\footnote{\url{https://se.mathworks.com/products/matlab-coder.html}, last accessed on 2020-12-15} we were able to generate C++ functions from the MATLAB code. For the simplest integration between the generated C++ functions and the JUCE plugin we chose to limit the function inputs and outputs to built-in and derived C++ data types. This required our MATLAB functions to have fixed-sized inputs and outputs. We decided on a maximum input/output size of 4096 double-precision floating point numbers, this being the maximum buffer size the plugin would be able to work with.  

A helper file was created to ensure code consistency, allowing the user and MATLAB coder to verify the functions with different inputs. Having this setup made it easy to go back to the MATLAB code and generate updated C++ functions without breaking the JUCE plugin.  

\subsubsection{TensorFlow in C++}\label{subsubsec:Tensorflow_in_C++}
Running the DDSP TensorFlow implementation in a real-time audio application is a heavy computational challenge. Moving from TensorFlow in Python to the TensorFlow C API\footnote{\url{https://www.tensorflow.org/install/lang_c}, last accessed on 2020-12-15} allowed us to integrate the models into the C++ codebase. By moving the TensorFlow computations to a separate thread, we can load the models, set the inputs, run the parameter estimation and save the outputs, without experiencing buffer underruns in the main audio processing thread.  

\subsubsection{Input signals}\label{subsubsec:input_signals}
The DDSP autoencoder needs the input values \emph{fundamental frequency} ($f_0$) and \emph{loudness} ($ld$). Since we allow both MIDI and line-in audio, two separate implementations are needed to calculate these values. Functions for this were created in MATLAB, but in the C++ implementation we chose to use the implementation of the YIN pitch tracking algorithm \cite{yin2002} from the C library Aubio\cite{aubio}, since it yielded more precise results.
%cite \cite{aubio} for aubio (pitch and loudness detection)

\subsection{Training models}\label{subsec:training_models}

\subsubsection{Pre-trained models}\label{subsubsec:pre-trained_models}
Next to the \emph{tone transfer} website mentioned in the introduction, the authors of the DDSP paper also published a Jupyter Notebook Demo on Google Colab called \emph{timbre transfer.}\footnote{\url{https://colab.research.google.com/github/magenta/ddsp/blob/master/ddsp/colab/demos/timbre_transfer.ipynb}, last accessed on 2020-12-15}
We accessed the available checkpoint files for violin, flute, tenor saxophone and trumpet from this notebook for our real-time implementation of the timbre transfer. However, we were not immediately able to use them in the JUCE plugin. The DDSP models are trained using TensorFlow's \emph{eager execution mode}, while the TensorFlow C API is constructed around \emph{graph mode}. Additionally, since we required the models to be controllable by MIDI input, we needed direct access to the decoder part of the model instead of supplying audio to the encoder. 

The \texttt{convert\_models.py} script from the Python folder of the plugin code repository deals with these requirements by loading the eager model from the downloaded checkpoint file, constructing a graph-based model only containing the decoder and then copying all weights from the old model to the new one. The resulting checkpoint now contains a graph that can be loaded by the TensorFlow C API.

\subsubsection{Custom models}\label{subsubsec:our_models}
In order to make use of the DDSP training library and extend the synthesizer with additional models, we created four custom models trained on:
\begin{itemize}
    \item Bass sounds of the Moog One, Moog Minimoog and Moog Minitaur synthesizers
    \item Studio recordings of Middle Eastern instruments, the Hammered Dulcimer and Santoor
    \item Studio recordings of a Handpan (also known as Hang Drum)
    \item Nature field recordings of birds chirping
\end{itemize}
For training we used the official DDSP (version 0.14.0) Jupyter notebook on Google Colab called \emph{train autoencoder}\footnote{\url{https://colab.research.google.com/github/magenta/ddsp/blob/master/ddsp/colab/demos/train_autoencoder.ipynb}, last accessed on 2020-12-15} which allows training on a Google Cloud GPU using own data. We chose the recordings listed above in order to obtain interesting sounds that differ from the more traditional pre-trained DDSP models. According to the recommendations of the DDSP authors given in the notebook, trained models perform best using recordings of a single, monophonic sound source, in one acoustic environment, in .wav or .mp3 format with a total duration of 10 to 20 minutes. Since the DDSP Autoencoder is conditioned on the loudness $A$ and the fundamental frequency $f_0$, i.e., the model learns to associate different synthesizer configurations to specific value pairs of $(A, f_0)$, training on multiple instruments, acoustic environments or polyphonic sounds prevents the autoencoder to learn a unified representation. However, these thereby introduced artifacts can also be used in a musical context, that is why we decided to challenge the autoencoder with less conform training data and eventually achieved interesting timbres. 

The training process is performed as follows. The first step is comprised of data generation and pre-processing of the training data. The raw audio is split into short parts of a few seconds, each analyzed on the specified features, i.e., the fundamental frequency and loudness, and finally saved in the TensorFlow \emph{TFRecord} format. The fundamental frequency is thereby estimated by using the state-of-the-art pitch tracking technique, called \emph{CREPE} by Kim et al. \cite{kim2018crepe} that applies a deep convolutional neural network on time-domain audio. The second step is the actual training, using a Python based configuration framework for dependency injection by Google, called \emph{Gin}\footnote{\url{https://github.com/google/gin-config}, last accessed on 2020-12-15}. In this way, all available training hyperparameters can be defined in a gin config file that is passed to the training function. The training process does not include any optimization techniques, such as a hyperparameter search or early stopping, the authors just recommend in the code documentation to train for 5,000 to 30,000 steps until a spectral loss of about 4.5-5 is reached for an optimal learning representation without overfitting. The third and last step is a short evaluation based on resynthesis. That means, a training sample is randomly picked, passed to the autoencoder that encodes and decodes, i.e., reconstructs the input sample based on the learned features.

We successfully conducted training of all four models and validated their performance in the previously mentioned timbre transfer demo. While validation using the DDSP library went smoothly and showed musically interesting results, we ran into issues during inference using the TensorFlow C API within our plugin. We monitored a much higher loudness of the custom models compared to the pre-trained models, resulting in a distorted, clipping sound. Furthermore, we detected a constant harmonic distribution independent of the incoming pitch and loudness while the pre-trained models adapt harmonics and frequency response according to these inputs. The overall experience with the training script provided by the DDSP authors is that it works without problems for standard parameters, but as soon as own hyperparameters within the gin framework are chosen, a lot of side-effects appear. For the mentioned reasons, integrating and possibly adapting the custom-trained models to make them work in the DDSP synthesizer will be a part of future work.

\subsubsection{Real-time implementation of the models}\label{subsubsec:realtime_implementation_of_the_models}
The DDSP non-real-time implementation synthesizes several frames before processing them into one output. Reading through the DDSP code base we experienced the number of frames (time steps) to be defined by the size of the input audio and a hop size defined by constants in the gin config file of the selected pre-trained model. 
\begin{comment}
\begin{equation}
   \text{hop\ size} = \frac{\text{number\ of\ samples\ in\ training}}{\text{number\ of\ time\ steps\ in\ training}} 
\label{eq:hop_size}
\end{equation}

\begin{equation}
    \text{time\ steps} = \frac{\text{size\ of\ audio\ input}}{\text{hop\ size}}
\label{eq:time_steps}
\end{equation}
\end{comment}

For our real-time implementation we wanted to calculate one frame with a size of the input buffer each time the buffer is ready. % Supporting a varying buffer size while following the equation for number of time steps \eqref{eq:time_steps} yields a new challenge. 
Given the static nature of our TensorFlow model implementation we were not able to change the number of time steps on the run. Therefore, we set the number of time steps to one. Each run of the TensorFlow model would then return a set of values for one time step, independent of the buffer size.  

\subsection{Additive synthesizer}\label{subsec:additive_synthesizer}
The implementation of the additive synthesizer can be found in the \texttt{additive.m} MATLAB code file. During the development of the DDSP synthesizer we went from a re-implementation of the DDSP equivalent to an adapted real-time optimized version with additional parameters for high-level control. While the original DDSP library provides two different implementations of the additive synthesis, the harmonic and sinusoidal approach, this work focuses on the harmonic synthesis that models a signal by adding only integer multiples of the fundamental frequency. 

In the following, the initial implementation as well as the main modifications in its final state are clarified.
As already explained in \ref{subsubsec:additive_synth}, the additive synthesizer models audio using a bank of harmonic sinusoidal oscillators. The synthesis algorithm takes amplitudes, harmonic distribution and fundamental frequencies for a specified number of frames as input and computes the sample-wise audio signal as output. The harmonic distribution provides frame-wise amplitudes of the harmonics. The additive synthesis as implemented in the DDSP library is performed in two main steps:
\begin{itemize}
    \item Translation of neural network outputs to the parameter space of the synthesizer controls
    \item Computing the output signal from synthesizer controls
\end{itemize}
In order to make the output of the neural network usable for controlling the synthesizer, it needs to be transformed accordingly. In detail, that means the amplitudes are scaled and the harmonic distribution is scaled, bandlimited and normalized while the fundamental frequencies remain unchanged. Bandlimiting the harmonic distribution means removing the harmonics that exceed Nyquist in order to avoid artifacts.

After retrieving valid synthesizer controls, the harmonic synthesis is performed. Since the DDSP approach works frame-based while the output needs to be delivered sample-based, the synthesizer controls need to be upsampled. This is done by linearly interpolating the frequency envelopes and windowing the amplitude envelopes by using 50\% overlapping Hann windows. Having calculated all controls on a sample basis, the signal can be synthesized by accumulative summation of the corresponding phases, i.e., adding the calculated sinusoids together, sample by sample. 

The following changes were made to optimize the algorithm for a real-time application and to add additional high-level control for the synthesis.
\begin{itemize}
    \item Since the frame-based calculation was computationally too heavy, we adapted the code so that the input is always one frame (equivalent to the buffer size) and all computations are sample-based. Therefore, no resampling or windowing is needed.
    \item Each time the function is called, the phases of all harmonics are saved and returned along with the signal and added as offset in the next call to avoid artifacts caused by phase jumps.
    \item In order to be able to optionally introduce non-harmonic partials to the signal, a stretch parameter was added that transforms the distance between the integer multiples while maintaining the fundamental frequency. An additional shift parameter adds the functionality to modify the fundamental frequency from one octave below to one octave above the current pitch in a continuous scale.
\end{itemize}

\subsection{Subtractive synthesizer}\label{subsec:matlab_subtractive_synth}
This component is responsible for the non-harmonic parts of instrument sounds, such as the audible non-pitched flow of air that accompanies the harmonic part of a flute sound. Our implementation, which can be found in the \texttt{subtractive.m} MATLAB code file, generates a frame of random noise and then filters it according to a given frequency response.

The function's parameters are the frame length (number of samples), noise color (see below) and the frequency response, which is given as a vector of $N$ magnitudes $m_0,\ldots,m_{N-1}$, where $m_0$ corresponds to the DC component and $m_i$ to frequency $f_{\text{nyquist}} / (N - i)$ with $f_{\text{nyquist}} = f_s / 2$ and samplerate $f_s$.

While we started with a direct re-implementation of the DDSP FilteredNoise approach described in \ref{subsubsec:ddsp_subtractive_synth}, we made the following adaptations over the course of the project:

\begin{itemize}
    \item Simplified filtering calculation. The DDSP synthesizer processes multiple frames at once. For the sake of a real-time implementation, we removed the step of calculating the impulse response for each frame and applying a windowing function. Instead, we simply perform a Fourier transform on the generated noise and multiply the result with the filter magnitude response that the model predicted for the single current frame.
    
    \item Noise color. We provide functionality to shape the frequency distribution of the generated noise. Noise color generally refers to the frequency $f$ being emphasized proportionally to $1/f^{\alpha}$ for some exponent $\alpha$ \cite{kasdin_1995}. $\alpha < 1$ results in higher frequencies becoming more prominent, while $\alpha > 1$ increases the energy of the lower frequencies. Uniform white noise is achieved by setting $\alpha = 1$. 
\end{itemize}

% \subsection{Reverb}\label{subsec:reverb}
% The DDSP library is doing reverberation using an impulse response created by the model. A real-time implementation of an impulse response reverb has not been implemented yet for this project. 

% \subsection{Modulation}\label{subsec:modulation}
% A modulation block has been planned for this project. By modulating a delay line and controlling the delay time, low frequency modulation rate and amount, the user will be able to add a simple chorus-or-delay type effect to the summed output of the additive and subtractive synthesizer. This module has not been implemented yet.  

\subsection{Graphical User Interface}\label{subsec:GUI}
After the development of all the features of our synthesizer, we focused our attention on designing an interface with high-level controls for the additive and the subtractive synthesis, the reverb, the modulation and the models. Our process started from a list of all the parameters we wanted to manipulate. We also looked for some inspiration from well-known VST synthesizers, comparing them in terms of usability and trying to understand what their best interaction features were. Later we organized the controls of our synthesizer in different modules and displayed them in a rectangular interface, trying to find a layout that was pleasant but also respectful of the instrument's architecture logic. In table \ref{tab:GUI_features}, we list all the controls for each module of our synthesizer. Because of the particular choice of a graphic control for the harmonics' amplitude, the team opted for a spectrogram representing the output of our plugin. In this way, the user is able to clearly see which harmonics are being played.

\begin{table}[htp]
    \begin{center}
    \begin{tabularx}{0.9\columnwidth}{|l|X|}
        \hline
        \textbf{Module}                             &   \textbf{Feature controls}\\
        \hline
        Input selector                              &   MIDI/line selector\\
        \hline
        \multirow{8}{50pt}{Models selector}         &   Violin\\
                                                    &   Flute\\
                                                    &   Saxophone\\
                                                    &   Trumpet\\
                                                    &   Moog Bass (not included)\\
                                                    &   Dulcimer (not included)\\
                                                    &   Handpan (not included)\\
                                                    &   Chirps (not included)\\
                                
        \hline
        \multirow{4}{50pt}{Additive synthesis}      &   Graphic harmonics editor\\
                                                    &   $f_0$ shift\\
                                                    &   Harmonics stretching\\
                                                    &   Global amplitude\\
        \hline
        \multirow{2}{50pt}{Subtractive synthesis}   &   Noise color\\
                                                    &   Global amplitude\\
        \hline
        \multirow{3}{50pt}{Modulation}              &   Modulation rate\\
                                                    &   Delay control\\
                                                    &   Amount\\
        \hline
        \multirow{3}{50pt}{Reverb}                  &   Dry/wet mix\\
                                                    &   Size\\
                                                    &   Glow\\
        \hline
        Output                                      &   Master gain\\
        \hline
        Spectrogram                                 &   Clear visualization of the output\\
        \hline
    \end{tabularx}
    \end{center}
    \caption{List of GUI's features}
    \label{tab:GUI_features}
\end{table}

Once we defined the layout and the parameters that we wanted to control, we moved to the software development in JUCE. In order to customize the appearance of knobs, we used the "Custom LookandFeel" objects while we designed ad hoc images for the buttons and background texture using a vector graphics software. Figure \ref{fig:ddsp_gui} previously presented the GUI of our synthesizer.

\subsection{Plugin setup}\label{subsec:plugin_setup}
The synthesizer ended up being built as a standalone executable and a DAW plugin using Steinberg’s VST3 format.  

Using JUCE’s AudioProcessorValueTreeState class we are exposing the different controllable parameters to the DAW, allowing control and automation of the plugin. Using this class we will also be able to easily store and read plugin states, enabling generation of presets, though this has not been implemented yet.

The synthesizer is configured to load the models from a given path with subfolders containing the individual models, as well as configuration files containing key-value pairs such as number of harmonics and scaling values.  

\section{Evaluation}\label{sec:evaluation}
In order to understand the strengths and weaknesses of our product to improve it, we designed an evaluation strategy for both User Experience (UX) and sound output. % We explained in section \ref{sec:design_requirements} that o
Our target users are musicians and music producers. Accordingly, we shared a release of our VST plugin with selected sound engineers, musicians and producers to collect opinions and user insights. Moreover, we designed two different questionnaires and asked participants to evaluate the UX and the sound accuracy of our software. The DDSP Synthesizer as well as the two questionnaires have been distributed online and the participants received an email with all the indications to properly conduct the test. 

In the next two sub-sections we will describe each evaluation in detail, including approach, desired outcome, survey design and results. 
% The UX survey is attached in appendix \ref{subcsec:ux_questions} and the sound accuracy survey is attached in appendix \ref{subsec:mushra_survey_content}. 
%Test your solution on target audience

\subsection{User Experience Evaluation}\label{subsec:ux_evaluation}
%report the right amount of time the participant had to evaluate the VST
\subsubsection{Approach}\label{subsubsec:ux_approach}
The aim of this evaluation is to collect feedback about the user interface from people with experience on synthesizers and music production. One of the goals of our project was to design a simple and efficient interface able to control several parameters with a single gesture without giving up functionality in the pursuit of simplicity.
After a trial period where the participants had the chance to familiarize themselves with the software, we asked them to compile a form.

\subsubsection{Survey structure}\label{subsubsec:ux_survey_structure}
Google Forms was chosen as a platform because of its simplicity and wide spread. We designed the survey with different sections to group the questions by theme. We included an experiment in order to ask each participant to load and perform some changes to a model and export the result in an audio file. In this way, we are sure that every participant had at least used and interacted with the plugin for a while. Moreover we are able to compare each audio export to understand if some of the instructions were not clear or if the UX itself was not effective. % The experiment details are reported in \ref{sec:appendix}. 

Four usage questions have been asked to collect information about the user's DAW and for how much time they used the plugin. In the next sections we asked the participants to report their experience during the experiment and evaluate the user interface rating 9 different statements with a Likert-scale, a widely used bipolar symmetric scaling method in questionnaires. In this way, users were able to express their agreement/disagreement related to each sentence.  Furthermore, we asked 4 open questions to let the participants express their opinion about the overall UX. Finally we added 8 questions to locate demographics and musical-related personal experiences. Table \ref{tab:ux_survey} summarizes the content of each section.

\begin{table}[htp]
    \begin{center}
        \begin{tabularx}{0.9\columnwidth}{|l|X|X|}
            \hline
            \textbf{\#} &   \textbf{Section}    &   \textbf{Content}\\
            \hline
            1           &   Introduction        &   Aim of the questionnaire\\ 
            \hline
            2           &   Experiment          &   Task instructions\\
            \hline
            3           &   Usage               &   4 mixed questions\\ 
            \hline
            4           &   UX evaluation       &   9 Likert scale evaluations\\
            \hline
            5           &   UX experience       &   4 open questions\\
            \hline
            5           &   Demographics        &   8 mixed questions\\ 
            \hline
        \end{tabularx}
    \end{center}
    \caption{Content of the UX survey}
 \label{tab:ux_survey}
\end{table}

\subsubsection{Expected results}\label{subsubsec:ux_expected_results}
In this section we want to express a projection of the feedback regarding the User Experience. Considering that the software is still under development, we are expecting reports about compatibility issues with different DAWs as well as some stability problems. Moreover, because of the VST's instability in the first release, it is possible that some users will not be able to conduct the small experiment that requires the plugin to be embedded in a DAW track. Considering the whole interface, one of the main points of our design requirements was the simplicity %   (\ref{subsec:research_question_statement} and \ref{subsubsec:documentation_of_requirements}) 
and thus our hope is to facilitate the user's interaction.

\subsubsection{Results}\label{subsubsec:ux_results}
We received 7 answers. Five participants identified as males, 1 female and one preferred not to say. The age average is 28.57 years (STD 8.42). Six of them declared that sound production is their hobby while one said music production is related to their job. The mean experience in the music production field is 7.43 years (STD 4.87). % and the music genres that participants reported working with is summarized in figure \ref{fig:ux_genres}. 
Six users do not have experience with machine learning VST plugins and only one of them does not know if she/he ever used one. Each user spent an average of 23.57 minutes using our synthesizer (STD 17.74). We suppose that some mistake has been made reporting the usage time for at least one user. In table \ref{tab:used_daw} we report the number of user tests per different software environment.

\begin{table}[htp]
    \begin{center}
        \begin{tabularx}{0.7\columnwidth}{|l|X|}
            \hline
            \textbf{\# users} &  \textbf{Environment}    \\
            \hline
            3   & Reaper\\
            \hline
            2   & Ableton Live\\
            \hline
            1   & Cubase\\
            \hline
            1   & Standalone version\\
            \hline
        \end{tabularx}
    \end{center}
    \caption{List of used DAWs in the evaluation.}
 \label{tab:used_daw}
\end{table}

\begin{figure*}[htp]
\centering
\includegraphics[width= 0.99\columnwidth]{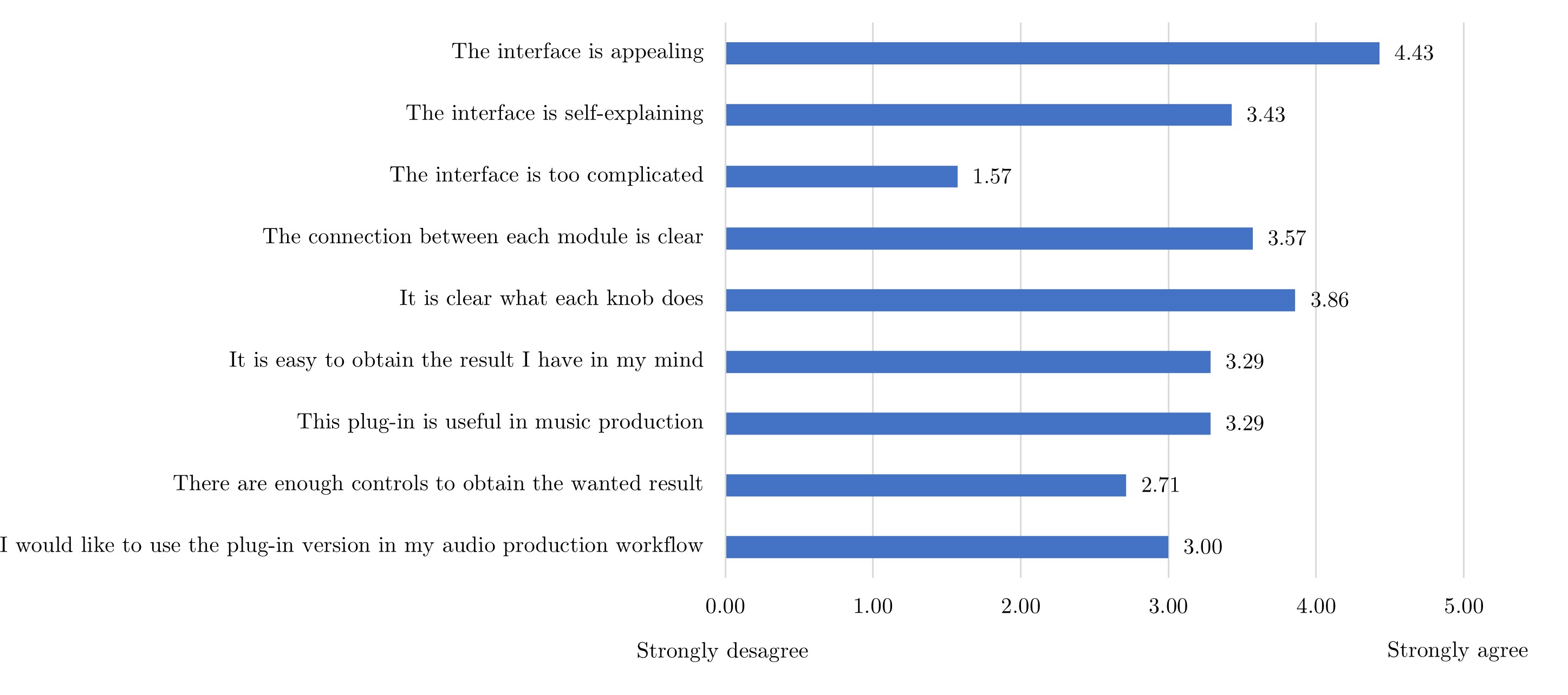}
\caption{User experience evaluation - Likert scale}
\label{fig:ux_likert_results}
\end{figure*}

In general, the experiment %explained in \ref{subsubsec:ux_survey_structure} and reported in appendix \ref{subsec:ux_experiment} 
has been rated a medium difficult task with a mean rating of 3.43 in a scale from 1 to 5 being 1 "easy to accomplish" and 5 "hard to accomplish". In figure \ref{fig:ux_likert_results} we summarize the answers obtained from the questions with an associated likert scale. The users were asked to rate each sentence from 1 to 5 with 1 corresponding to "strongly disagree" and 5 to "strongly agree". We can observe that the graphical user interface has been really appreciated with a 4.43 mean value while the interface's controls seem not to let the participants easily reach the wanted results. The other statements reported in the likert section obtained a medium rating between 3 and 3.86 which might mean that the GUI is in general appreciated.

As expected, some of the participants encountered difficulties in the installation procedure of the VST3 plugin in both Windows and macOS environments while the standalone version seems to be more stable. Furthermore, three users reported an unsatisfactory audio result related to the presets obtained from models. Here we report part of one of the feedback: "\textit{[...] It's possible to get some cool sounds but the default sound when you just start it is not so nice.}". On the other hand, the audio input feature was appreciated: "\textit{[...] I think the audio input feature has a lot of potential and I caught myself experimenting with this a lot and loosing track of time.}". Two participants reported that the possible interaction with the interface for the additive synthesizer was not immediate to spot and they realized its features after a while. For this reason they suggest a graphical indication to guide the user to the interaction with the harmonic sliders.
A significant outcome is the unexpected audio results that participants reported. Even though they described output sounds as "awkward", they highlighted the new creative way of producing unexpected sounds, finding the whole synthesizer experience engaging.

%%%%%%%%%%%%%%%%%%%%%%%%%%%%%%%%%%%%%%%%%%%%%%%%%%%%%%%%%%%%%%%%%%%%%%%%%%%%%%%%%%%%%%%%

\subsection{Design Requirements Validation}\label{subsec:designrequirements_validation}

To validate if our synthesizer implementation met our design requirements we conducted two evaluations, which were presented in the previous sections. 
Design requirement 1: \emph{‘Must work in real-time as a playable software instrument’} and 2: \emph{‘Must be implemented as a software plugin’} has been met as our software synthesizer plugin features models that are implemented to be playable in real-time (elaborated in \ref{subsubsec:realtime_implementation_of_the_models}). We confirmed that these requirements were met by running user-tests where participants had to install the VST3 plugin as well as play it as an instrument in real-time \ref{subsubsec:ux_results}. Design requirement 3: \emph{‘Must allow switching between Line and MIDI input’} and 4: \emph{‘Must implement at least 4 pre-trained models’} was met and addressed in the user-tests as well. Participants reported that switching between Line and MIDI input makes sense, and that they especially enjoy the Line-input functionality \ref{subsubsec:ux_results}. The four pre-trained models that we implemented in our synthesizer were violin, flute, saxophone and trumpet. The timbre transfer capabilities of these models in context of our implementation, were evaluated in section \ref{subsubsec:mushra_results}. Furthermore, we trained additional models but those were not included in this version of the synthesizer as stated in section \ref{subsubsec:our_models}. Lastly, the fifth design requirement \emph{‘Must include graphical UI components for intuitive manipulation of synth parameters’} was partially met with our implementation of the individual partials control GUI, however as reported in section \ref{subsubsec:ux_results} the participants found the general UI to be visually appealing and self-explanatory, but the graphical interface for controlling the harmonic distribution can be improved by making its functionality more obvious to the user. Additionally, implementation of the "modulation" section of our synthesizer would contribute to fulfilling this design requirement as participants reported a lack of controls to achieve a desired output sound \ref{subsubsec:ux_results}. The current state of our synthesizer sufficiently meets two of our design requirements while partially meeting the other three.

\section{Discussion}\label{sec:discussion}

% This section aims to discuss some of the challenges we encountered during the implementation as well as the results of the evaluation concerning different areas of the project. In each area of the project, we will address the insights gained and offer a roadmap for future improvement.

\subsection{Synthesizer controls}
From the Likert evaluation (see figure \ref{fig:ux_likert_results}), we can conclude that users found the interface to be visually appealing and for the most part intuitive to understand, with the exception of the graphical control over the harmonic distribution. However, there is also an indication that the users desired more control over the sound and did not find themselves being able to reach the expression they were trying to achieve. We attribute this sentiment partially to some of the initially planned modules not being implemented in the distributed product. To address this, in the future we would like to fully implement the reverb and modulation modules, and support MIDI Polyphonic Expression (MPE) features.
\begin{comment}
\begin{itemize}
    \item The reverb module and its user controls
    \item The modulator and the ability to connect it to both the additive and the subtractive synthesizer
    \item Controls that allow the user to shape the magnitude response of the filter used in the subtractive synthesizer
    \item Additional explanation and a way to reset the harmonic distribution interface
    \item Support for MIDI Polyphonic Expression (MPE) features such as Aftertouch
\end{itemize}
\end{comment}

\subsection{Real-time timbre transfer}

% The webMUSHRA evaluation (see figure \ref{fig:mushra_resultsbar}) as well as our own subjective impression place 

We found the quality of the timbre transfer in our real-time implementation below that of the demonstrations published by the Magenta team. Our converted models preserve some characteristics of the original ones, such as wind noises in the flute model, but do not accurately reproduce the timbre overall. We confirmed that on the level of a single frame, our models produce the same output as their original counterparts; will investigate and improve the quality in the future. Additionally, we would like to further investigate why we were unable to perform the timbre transfer with models that we trained both within the framework provided by Magenta, and within custom environments.

A recently released realtime reimplementation of DDSP in PyTorch\footnote{\url{https://github.com/acids-ircam/ddsp_pytorch}} provides a possibly more seamless way of interfacing with DDSP models in C++ that proved compatible with our plugin and JUCE. Extending that API to allow the user some control over the synthesis parameters seems a promising avenue to improve the sound quality of our timbre transfer.

\begin{comment}
so we propose the following potential reasons for the quality mismatch that we would like to 

\begin{itemize}
    \item Frame-based tracking of the input. Where the timbre transfer demo uses a deep convolutional neural network that operates on the entire input, our project makes use of the Aubio implementation of the YIN algorithm \cite{yin2002}, considering only up to 4096 samples at a time to track the input audio and convert it into f0 and loudness data. This leads to a loss of information and accuracy when it comes to pitch and note onset detection, and it would stand to reason that less accurate input data leads to less accurate output from the models.   
    \item The models were originally trained on receiving the loudness and pitch information for seconds of audio at a time, and generate the synthesizer controls using a recurrent neural network, meaning that future output depends on previous inputs to a degree. We suspect that a lot of that information is lost through our implementation decision to generate the audio independently from frame to frame. We plan to experiment with keeping internal buffers of previous input to remedy this, keeping an eye of the impact larger input sizes to the models might have on real-time performance. 

\end{itemize}
\end{comment}

\subsection{Distribution as a VST3 plugin}

When it came to distributing our project to users, we encountered some difficulties in packaging the required libraries and model files together with the generated VST3 plugin. Some of the DAWs that users tested on, like Ableton or Reaper, did not recognize the plugin or experienced stability issues during its usage. 
Although the core functionality could still be accessed via the standalone application generated by JUCE, the project was designed first and foremost as a plugin. Functionality like handling of external audio sources and wet/dry mixing was expected to be handled by the host DAW. Users who had to resort to the standalone when their DAW did not recognize or stably run the plugin reported those features as missing.

Thus, we would like to improve the distribution process in the future, ensuring that the project can be seamlessly installed as a plugin in multiple DAWs on Windows and macOS.

\section{Conclusion}\label{sec:conclusion}

In this paper, we presented an approach to integrate the DDSP library into a real-time plugin and standalone application using the JUCE framework. We succeeded in implementing a synthesizer playable based on pure user input. While we were generally able to use the output from pre-trained models to control the DDSP backend, further research is needed to match the sound quality of these real-time models to that of the offline timbre transfer examples provided by the DDSP authors.

\bibliography{SMC704_DDSP_Library}

\begin{thebibliography}{10}

\bibitem{donahue_adversarial_2019}
C.~Donahue, J.~McAuley, and M.~Puckette, ``Adversarial {Audio} {Synthesis},''
  {\em arXiv:1802.04208 [cs]}, Feb. 2019.
\newblock arXiv: 1802.04208.

\bibitem{Blaauw2017}
M.~Blaauw and J.~Bonada, ``A neural parametric singing synthesizer modeling
  timbre and expression from natural songs,'' {\em Applied Sciences}, vol.~7,
  p.~1313, 2017.

\bibitem{engel_ddsp_2020}
J.~Engel, L.~Hantrakul, C.~Gu, and A.~Roberts, ``{DDSP: Differentiable Digital
  Signal Processing},'' {\em International Conference on Learning
  Representations}, 2020.

\bibitem{serra_spectral_1990}
X.~Serra and J.~O. Smith, ``{Spectral modeling synthesis. A sound
  analysis/synthesis system based on a deterministic plus stochastic
  decomposition},'' {\em Computer Music Journal}, vol.~14, no.~4, pp.~12--24,
  1990.

\bibitem{native-instruments}
Native-Instruments, ``Razor,'' 2011.

\bibitem{Pandey2010}
D.~Pandey, U.~Suman, and A.~Ramani, ``An effective requirement engineering
  process model for software development and requirements management,'' in {\em
  2010 International Conference on Advances in Recent Technologies in
  Communication and Computing}, pp.~287 -- 291, 2010.

\bibitem{yin2002}
A.~Cheveigné and H.~Kawahara, ``{YIN, A fundamental frequency estimator for
  speech and music},'' {\em The Journal of the Acoustical Society of America},
  vol.~111, pp.~1917--30, 2002.

\bibitem{aubio}
P.~Brossier, {\em Automatic annotation of musical audio for interactive
  applications}.
\newblock PhD thesis, Queen Mary University of London, 2006.

\bibitem{kim2018crepe}
J.~W. Kim, J.~Salamon, P.~Li, and J.~P. Bello, ``{CREPE: A convolutional
  representation for pitch estimation},'' in {\em 2018 IEEE International
  Conference on Acoustics, Speech and Signal Processing (ICASSP)},
  pp.~161--165, IEEE, 2018.

\bibitem{kasdin_1995}
N.~J. Kasdin, ``Discrete simulation of colored noise and stochastic processes
  and $1/f^\alpha$ power law noise generation,'' {\em Proceedings of the IEEE},
  vol.~83, no.~5, pp.~802--827, 1995.

\end{thebibliography}

\end{document}